
\input harvmac

\def\pt{\partial}

\def\CZ{{\cal Z}}
\def\CS{{\cal S}}

\def\l{\ell}

\def\CW{{\cal W}}

\def\p{\partial}
\def\CK{{\cal K}}

\def\CD{{\cal D}}
\def\R{\relax{\rm I\kern-.18em R}}
\font\cmss=cmss10 \font\cmsss=cmss10 at 7pt
\def\Z{\relax\ifmmode\mathchoice
{\hbox{\cmss Z\kern-.4em Z}}{\hbox{\cmss Z\kern-.4em Z}}
{\lower.9pt\hbox{\cmsss Z\kern-.4em Z}}
{\lower1.2pt\hbox{\cmsss Z\kern-.4em Z}}\else{\cmss Z\kern-.4em Z}\fi}
\def\pl{{\it  Phys. Lett.}}

\def\np{{\it Nucl. Phys. }}

\def\CP{{\cal P}}
\def\s{\sigma}

\def\pt{\p_{\tau}}

\def\CJ{{\cal J}}

\lref\ackm{J. Ambjorn, L. Chekhov, C. Kristjansen and Yu. Makeenko, \np B
404 (1993) 127; \ J. Ambjorn and C. Kristjansen, {\it Mod. Phys. Lett. } A 8
(1993) 2875}
 \lref\KlSu{I. Klebanov and L. Susskind, \np B309 (1988) 175}
\lref\gk{  D. Gross and I. Klebanov, \np B344  (1990)
475; G. Parisi, \pl 238 B (1990) 213 }
  \lref\ms{G. Moore and N. Seiberg, {\it Int. J. Mod. Phys.} A 7 (1992) 2601}
 \lref\moor{G. Moore, \np B 368 (1992) 557}
\lref\Imat{I. Kostov and M. Staudacher, \np B  384 (1992) 459}
 \lref\Iade{I. Kostov, \np B 326, (1989) 583}
\lref\adem{I. Kostov, \pl    297 B (1992)74}
 \lref\mss{G.Moore,
N. Seiberg and M. Staudacher, \np B 362 (1991) 665}
   \lref\ksks{I. Kostov and M. Staudacher,  \pl B 305 (1993) 43}
 \lref\kko{V. Kazakov and I. Kostov, \np B 386 (1992) 520}
 \lref\djv{  S. Das and A.Jevicki,
{\it Mod. Phys. Lett.} A 5 (1990) 1639}
 \lref\Idis{I.K. Kostov,  \np B 376 (1992) 539}
 \lref\IM{I. Kostov and M. Staudacher, \pl B 305 (1993) 43}
  \lref\polch{J. Polchinski \np B 362 (1991) 125}
\lref\aff{I. Affleck,Nucl. Phys. B185 (1981) 346}
\lref\miko{M. Douglas, Collective field theory  of open and closed strings
in 1+1 dimensions, not published}
\lref\min{ J. Minahan, preprint UVA-HET-92-01, March 1992}
\lref\berk{M. Bershadsky and D. Kutasov, \pl B 274 (1992) 331 }
\Title{}{\vbox{\centerline{
    Field Theory of Open and Closed  Strings}
\vskip4pt\centerline{
   with Discrete Target Space }
}}
 \bigskip\centerline{ I. K. Kostov \footnote{$^{\ast } $}
{ on leave from the Institute for Nuclear Research and Nuclear Energy,
72 Boulevard Tsarigradsko Chauss\'ee, 1784 Sofia,
Bulgaria}\footnote{$^{\diamond}$}{e-mail:kostov@amoco.saclay.cea.fr}}
\bigskip\centerline{{\it Service de
 Physique Th\'eorique}  \footnote{$^{\dagger}$}{
Laboratoire de la Direction
 des Sciences de la Mati\`ere du
Commissariat \`a l'Energie Atomique}{\it  de Saclay} }
\centerline{{\it CE-Saclay, F-91191 Gif-sur-Yvette, France}}

\vskip .3in

\baselineskip8pt{ We study  a $U(N)$-invariant vector+matrix chain with the
color structure of a lattice gauge theory with quarks and interpret it as a
theory of open and
closed strings with target space $\Z$. The  string field theory is constructed
as  a quasiclassical
expansion for the Wilson loops and lines in this model.  In a particular
parametrization this is  a theory  of
two scalar massless fields defined in the half-space $\{x\in \Z , \tau >0\} $.
The extra
dimension $\tau$ is related to the longitudinal mode of the strings. The
topology-changing  string
interactions are described by a local potential. The  closed string interaction
is nonzero only at  boundary $\tau =0$ while the open string interaction falls
exponentially with $\tau$.
 }

\bigskip
\bigskip
\bigskip
\leftline{Submitted for publication to: {\sl   Physics Letters B}}%
\rightline{ SPhT/94-097}
\Date{10/94}
\baselineskip=20pt plus 2pt minus 2pt

\bigbreak\bigskip\bigskip\centerline{{\bf Introduction}}\nobreak

The $D=1$ closed string theory is known to  describe a special critical
behaviour of one-dimensional $N\times N$ matrix models. Remarkably, the
discrete matrix chain   leads to the same string theory as the  continuum model
(known as matrix quantum mechanics), under the condition that   the lattice
spacing  $\Delta$ is smaller than some critical value, which we assume equal to
1. At
 $\Delta=1$ the system undergoes a  Kosterlitz-Thouless   type transition  and
if  $\Delta>1$ the matrices  decouple \gk .

 The appearance of a minimal length in the target space, anticipated by
Klebanov and Susskind in \KlSu , seems to be a fundamental property of the
string theory.  It signifies  that the string theory has much fewer
short-distance degrees of freedom than the conventional quantum field theory.
As a consequence of this,
the continuous target space can be
restricted to  a lattice $\Z \subset \R$    without loss of
information\foot{It has been checked   \IM\ that  the $n$-loop tree-level
amplitudes ($n\le 4$) in the   closed string theory with target space $\R$ can
be reproduced   from their restrictions in $\Z$.}.

 The physics in the  target space $\Z$  seems to be simpler than in $\R$. The
loop amplitudes restricted to  $\Z$ enjoy some nice factorization
properties.   Furthermore, as a consequence of the periodicity of the momentum
space,  an infinite set of "discrete" states  with  integer momenta  become
invisible in the space $\Z$.   Therefore, it seems  advantageous to consider a
string theory  on a lattice with   spacing  not inferiour but equal to   the
Kosterlitz-Thouless distance.

The string theory with   target space $\Z$    has been originally constructed
in  \Iade\  as an SOS model on a surface with fluctuating geometry. The secret
of its solvability is the
possibility to be   mapped onto a gas of nonintersecting (but otherwise
noninteracting)  loops on the
world sheet.   The loops on the world sheet define a natural
discretization of the moduli space and  a possibility to construct an
unambiguous   string field
diagram technique \Idis.   The loop gas approach can be easily generalized to
to    open strings.  The critical behavior of the  open   strings with target
space $\Z$  has been
studied  by the loop gas
method in \kko .

To  study of  the topology-changing interactions of
open and closed strings  we need   more advanced technology   than the
world sheet surgery  applied in \Idis\ and   \kko  . Such might be provided by
 an underlying  large $N$ field-theoretical model, as in the case of the
continuum string theory.

 In this letter  we  construct  a one-dimensional lattice model with local
$U(N)$ symmetry, whose  color structure is that of a lattice gauge theory with
quarks, and  show that it is equivalent to the field theory of closed and open
strings in $\Z$.
  The mean field problem in this model  is the one-matrix integral with
potential determined dynamically.  The  quasiclassical  expansion  for the
Wilson loops and lines   yields
  the string field Feynman rules.
 The vertices for the string fields have the geometrical interpretation of
surfaces with various
topologies localized at a single point of the target space.

In a special parametrization, the effective
action is this of    theory of scalar fields in the comb-like space $(x,\tau),
\ x\in\Z, \tau >0$.
  The kinetic term for   these fields  involves  finite-difference operators in
$x$ and $i\tau$
directions. The   interactions   are described by a local nonpolynolial
potential.
 The   closed strings interact only along the boundary $\tau =0$ while  the
coupling of the
open strings   falls exponentially in the bulk.

\bigbreak\bigskip\bigskip\centerline{{\bf Closed and
open strings from a $U(N)$ matrix-vector chain}}\nobreak

   The underlying lattice model
possesses local $U(N)$  symmetry and resembles a  Wilson lattice gauge theory,
with  the  unitary
measure for the "gluon" field replaced by a  Gaussian measure.     The Gaussian
measure allows the eigenvalues of the gauge field to fluctuate in the radial
direction, which  leads  to the longitudinal (Liouville)
mode of the string.    The  vacuum energy of the model is  equal to the
partition function of a gas of triangulated surfaces with free boundaries,
immersed in the lattice $\Z$.

The fluctuating  variables associated with each point $x\in \Z$ are a
 fermion
   $\psi_x= \{ \psi_x^i\}, \bar \psi_x= \{\bar \psi_x^i\}$, a
hermitian matrix  $\Phi_x=
\Phi_x^{\dag}=\{ \Phi_x^{ij} \}   $, and a complex matrix $A_x =
\{A_{x}^{ij}\}$  with color indices
ranging from 1 to $N$.

  To simplify notations we combine the  color index $i$ and the space
coordinate $x$ into a   double index  $  a=\{i,x\},\ \ \ i=1,...,N, \ \ \ x\in
\Z$.
 Then the entities of the model are the   vector  with  anticommuting
coordinates
\eqn\vvec{\psi_{a} =\psi_{ix}, \ \ \ \bar \psi_{a}  =\bar\psi_{ix} , }
 and the hermitian matrix
\eqn\xmm{ A_{a  a'}  =  \delta_{x,x'}\Phi_x ^{i j}
+\delta _{x,x'-1} \  A_{ x   }^{i  j} +\delta _{x,x'+1} \  A^{\dag \ i j}_{ x
-1}  ;
\ \ \ \ \  a=\{ i,x \} ,\
a'=\{j, x'\}. }
The partition function  is defined  by   the    integral
\eqn\partf{\CZ=\int [ dA ] \ d\bar \psi d\psi\exp\big[ - {1\over 2}   \tr
A^{2}
+{\lambda \over 3\sqrt{N}}  \tr
 A^{3}  -    \bar \psi  \psi + \lambda_{B} \bar \psi  A\psi\big]}
 where the trace is understood in the sense of the double index $a$ and
$[dA]$ is the homogeneous
  measure for the nonzero matrix elements \xmm .

The perturbation series for
the  free energy  $\CF= \log \CZ$  is a sum over connected "fat" graphs  dual
to triangulated
 surfaces with  boundaries.
The "windows" of the fat graph  are spanned on the index lines labeled by
double indices
$a=\{ i, x\}$. Therefore  an integer
coordinate $x$ is assigned to each point of the  triangulated surface.
   The free energy is   equal to the sum of all connected surfaces $\CS$ with
free boundaries,
immersed in $\Z$
 \eqn\surs{\CF= \sum  _{\CS}(- N)^{\chi} \lambda^{S} \lambda_B^{L_B} }
where  $\chi = \ 2 \ - 2  \# {\rm (handles)}
 -  \# {\rm (boundaries)} $ is the Euler characteristics,   $S=  \#$(triangles)
 is the area,
  and  $L_B = \# $(edges)  is the total length of the boundaries  of the
surface
$\CS$.  The gauge invariant operators creating closed and open strings are  the
  Wilson loops
and lines constructed in the same way as in the lattice gauge theory. We
restrict ourselves to closed and open strings  localized at
 a single point $x $
\eqn\wllp{W_{x}(\l)= \tr \ e^{\l \Phi_x } , \ \ \ \Omega_{x}(\l) = \bar \psi
_{x }   e^{\l \Phi_x }
\psi_x  }
where the parameter $\l$ is the (intrinsic) length of the string. Since the
time slice of the one-dimensional spacetime consists of a single point, the
operators \wllp\  generate the whole Hilbert space.
In this case the $A$-matrices are redundant variables and will be integrated
out.

In the following
we will   consider  a more general action containing    source terms  $J$ and
$J^B$. It is convenient to absorb  the  coupling
constants $\lambda$ and $\lambda_B$  into the source    and    shift   $\Phi_x
\to    \Phi_x +(2\lambda)^{-1} I$, where $I$ is the unit matrix.  Then,   after
performing the
Gaussian integral over the  $A$-variables,  we find
\eqn\mtra{\CZ[J,J^B]=\int \prod _{x}\ d\psi_x e^{\tr\ J_{x}(\Phi_{x})}
\ d\bar \psi_x d\Phi_x  e^{  \bar  \psi_x \ J_{x}^{B}(\Phi_{x}) \psi_x } \ \
e^{\CW }}
\eqn\mtrb{\eqalign{  \CW&=
    - \half \sum_{x,x'}  C_{xx'} \Big(    \log | \det
(I \otimes \Phi_x +  \Phi_{x'}\otimes I)|   \cr
& + [\psi_x \otimes \bar \psi _{x'}]
[ I \otimes \Phi_x +  \Phi_{x'}\otimes I ]^{-1}  [\psi_{x'} \otimes \bar\psi
_x]\Big)  \cr}}
where  by $C_{x x'}$ we denoted the  incidence  matrix
 of the target space lattice $\Z$
 \eqn\ccmc{C_{x x'}= \delta _{x, x'+1}+\delta _{x, x'-1}.}

As a consequence of the local $U(N)$      symmetry  the only relevant degrees
of freedom are the
$N$ real eigenvalues $\phi_{ix}  $ of the hermitian matrix $\Phi_x$ and the
 commuting   nilpotent  variables  $\theta_{ix}  =  \psi_{ix}
 \bar \psi_{ix}$.
The integration measure factorizes into the Haar measure in the $U(N)$ group
and an integration measure along the radial directions $\phi_{ix}, \theta_{ix}$
\eqn\intm{ d\Phi_{x} \ d\bar\psi_xd\psi_x = {\rm constant} \times \prod
_{i=1}^N
d\phi_{ix}d\theta_{ix} \  \Delta^2 (\phi_x) }
where  $\Delta(\phi)$ is   the Vandermonde determinant
\eqn\wand{\Delta(\phi)=\prod _{i<j} (\phi_i -\phi_j).}
  The algebra  and the integration over the  $ \theta  $-variables  are defined
by the rules
 \eqn\grrr{  \theta\theta ' =\theta '\theta , \ \ \theta^2=0 , \ \   \int d
\theta  =0 , \  \int d\theta \ \theta =1.}
 To save space we  will use the following compact notations
\eqn\cnts{   \hat \phi_{ix}=
\{  \phi_{ix} ,  \theta_{ix} \} , \ \   d \hat \phi_{ix}=d\phi_{ix}\
d\theta_{ix},}
\eqn\resr{  \hat   J_x(\hat \phi_{ix}  )  = J_x(\phi_{ix})+     \theta_{ix}
J_x^B(\phi_{ix}). }
The partition function \mtra\  reads, in terms of the radial variables
$\hat\phi_{ix}$,
 \eqn\mtria{
\CZ[\hat J ] =     \int \prod _{x}
   d^N\hat \phi_{ x}  e^{\hat J_x (\hat \phi_{ix}) } \  \Delta (\phi_x)
\  e^{
 \CW [\hat \phi ]} }
\eqn\mtrin{  \CW  [\hat \phi ]  =   - \half \sum_{x, x'; i, j}  C_{xx'}
  \ln  |\phi_{ix}+\phi_{jx'}+\theta_{ix}\theta_{jx'}    |
        }

 The partition function \mtria\ generalizes the  eigenvalue integral for the
pure matrix theory
introduced  in \adem .
 Note that the   pure matrix theory (no $\theta$'s)  describing the  closed
string sector
 can be  reformulated, using the Cauchy identity
\eqn\iiz{ {\Delta(\phi ) \Delta(\phi ') \over \prod
_{i,j} (\phi_{i } +\phi '_ j)}= \det {1\over \phi_{i} +\phi'_j},}
as a free Fermi system defined by a one-particle transfer matrix, much  as
the matrix quantum mechanics.
However, after introducing the "quark" fields,   the
fermions of different colors start to interact. This is the main obstacle for
generalizing the
   formalism of matrix quantum mechanics to open strings\foot{This problem was
  have been considered originally by I. Affleck \aff \ and, more recently,  by
J. Minahan \min \ and M.
Douglas\miko .}.

\bigbreak\bigskip\bigskip\centerline{{\bf  A field theory for the loop
variables}}\nobreak

 The density $ \hat \rho _x =\s_x  +
\theta \rho_x  \  $ for the distribution of the radial
coordinates $\hat \phi_{ix}= \{  \phi_{ix} ,  \theta_{ix} \}$
  \eqn\denns{\eqalign{
    \hat \rho _x(  \phi , \theta  )  & = \delta (\phi - \Phi_x)\ \delta (\theta
-       \bar \psi_x \psi_x ) =  \sum _{i=1}^N
\delta(\hat \phi , \hat \phi_{ix} )\cr
&= \sum _{i=1}^N (\theta + \theta_{ix}) \
\delta( \phi -  \phi_{ix} ) =\s_x(\phi) +
\theta \rho_x(\phi) .     \cr}}
  is the collective field  for which the
$1/N$ expansion makes sense of  quasiclassical expansion.
 The Laplace transform of the density  ($ \varepsilon $ is assumed to be  a
nilpotent variable as $\theta$)
 \eqn\tfdr{ \hat W_{x}(\l , \varepsilon) =  W_x(\l)+ \varepsilon  \Omega_x(\l)
= \int  d\hat \phi \ e^{ \l\phi +\varepsilon\theta  }
\hat \rho_x(\phi , \theta).}
   gives the   Wilson loop  and line  \wllp .

 Let us perform a
 change of  variables  $\hat \phi_{ix}
  \ \to\  \hat \rho_{x}(\hat \phi ) $ in  the integral  \mtria .
The action $\CW$ becomes  a quadratic form in $\hat\rho$
 \eqn\kke{W [\ \hat\rho \ ]=   \half \hat\rho \cdot \hat K \cdot\hat\rho   =
\half \rho \cdot  K  ^+ \cdot\rho +\half  \s \cdot K^B \cdot\s  }
 where  $\ \cdot\ $  stands for a sum and integral over the repeated variables
 and the kernel $\hat K$ reads explicitly
 \eqn\krkrr{
  \hat K_{xx'}(\hat\phi , \hat\phi ')  =- C_{xx'}\ln |\phi +\phi '
+\theta\theta'|
=    K_{xx'}^{+} ( \phi ,  \phi ') + \theta \theta ' K_{xx'}^B ( \phi ,  \phi
') }
\eqn\kerns{ K_{xx'}^{+} ( \phi ,  \phi ')= -C_{xx'}\ln |\phi +\phi '|, \ \
K_{xx'}^B ( \phi ,  \phi ')= -{C_{xx'}\over  \phi +\phi ' }.}

  The  Jacobian
is expressed as usually as a functional integral over  a Lagrange multiplier
field\foot{In ref.
 \adem\ we have
included  the Vandermond  determinant
 $\Delta(\phi_x)$   in the effective action. The  resulting    collective
theory  is not well defined at short distances and therefore  ambiguous beyond
the tree level.
  The healthy   way to go to loop variables is to include the  Vandermond
determinant
into the Jacobian $\CJ[\rho]$.}
$\hat\alpha_x(\hat\phi) = \alpha_x(\phi) + \theta \beta_x(\phi),$
\eqn\jacc{\eqalign{ \CJ [ \hat\rho ]&=
   \prod_{x}  d^N  \hat \phi_{ x}     \  \Delta (\phi_{ x})
 \int  \CD \hat \alpha \
  \exp  \Big[  -\int d\hat \phi
       \hat   \alpha_x(\hat \phi)
     \big[  \hat\rho_x(\hat \phi)
 -  \sum_{i=1}^N \delta (\hat \phi -\hat \phi_{ix}) \big]
  \big) \Big] \cr
&=   \int  \CD \hat \alpha_x    e^{ -\hat\alpha_x \cdot \hat  \rho_x +
    \CF_0 [  \alpha_x +\ln \beta_x] }   \cr
 }}
where by $  \CF_0[V]$ we denoted  the logarithm of the one-site integral in
external field
$-V(\phi)$
\eqn\onmm{  e^{ \CF_0[V ]}=    \int \prod_{i=1}^N  d \phi_i\  e^{
V(\phi_i)} \
 \Delta^2 (\phi)\ . }

   Combining \kke\    and \jacc\  and  shifting
 $\alpha \to \alpha - \ln \beta +J, \beta\to\beta+J_B$,
 we write the partition function \mtria\ as  the following functional integral
 \eqn\ffpp{ \CZ [\hat J] =  \int \CD \hat\rho  \CD \hat \alpha \
\CW _{\rm tot}[\hat\rho,  \hat \alpha ] }
 \eqn\fpfp{\CW_{\rm tot}[\hat\rho,  \hat \alpha ]=
 \half \hat \rho \cdot \hat K \cdot\hat \rho   -
\hat\rho\cdot \hat \alpha
  +\sum_x \Big(  \CF_0[\alpha _x +J_x]   +    \int d\phi   \rho_x (\phi)  \ln
(\beta_x  +
J^B_x) \Big).}

The string field theory will be obtained as the large-$N$
 quasiclassical  expansion for this integral.
 For this purpose we have to solve the following technical problems: 1) find
the classical
string background  $\hat\rho_c, \hat\alpha_c$, which is the solution of the
saddle-point equations,
2) diagonalize the quadratic action, and 3) expand the  interacting
part\foot{We will treat the
$1/N$-corrections to the tadpoles and propagators as interaction} as a series
in $1/N ,
  \hat\rho -  \hat\rho_c, \hat\alpha-  \hat\alpha_c$.
The solution of the first two problems is known (see refs.
\Idis  ,\kko),
 and we will explain it without going  into details.

\bigbreak\bigskip\bigskip\centerline{{\bf  Saddle point}}\nobreak

     The
stationarity condition  for the $\hat \alpha$-field     gives
 \eqn\sdpt{\rho_c=  \Bigg({\delta  \over \delta \alpha } \CF_0[\alpha  +J
]\Bigg)_{\alpha=\alpha_c} , \ \
\s_c = {\rho_c  \over \beta_c+J^B}}
The first equation means that $\rho_c$ coincides with the classical  spectral
density in the
one-matrix   integral, which   is related
to the potential $V=-\alpha_c$ by  a linear equation.  If we denote by $ K^- $
the linear operator with kernel ($\CP$ means principal value prescription)
\eqn\yadr{K^-_{xx'}(\phi, \phi ') = -2\delta_{xx'} \CP \ln |\phi -\phi '|,}
then the first eq. \sdpt\  takes  the form
 \eqn\uedno{ K^- \cdot  \rho_c =\alpha_c. }
   Taking into account \sdpt , \uedno\ and neglecting the subleading   term
$\ln\beta$, we write the
stationarity conditions for the $\hat\rho$-field as
\eqn\trok{  (  K^- - K^+  )\cdot \rho_c\  = \   J  ,\ \ \ \ \
  {\rho_c \over \s_c }  -K^B \cdot  \s_c \  =\ J^B  .}

The first equation \trok\ determines the closed string background. It has been
solves exactly for
  a stationary polynomial source $J$
  \Imat , \ref\gaud{M. Gaudin,
unpublished} . In the scaling limit $\{   \lambda\to \lambda^* , N\to\infty
; \ N(\lambda^*-\lambda)=\Lambda \}$  the
solution is supported by a semi-infinite interval
 \eqn\pish{-\infty <\phi<-\sqrt \mu    }
 and reads explicitly
\eqn\cozh{\rho_c(\phi)=   \sqrt{\phi^2 -\mu} , \ \ \  (   1<<\mu<<N).}
The  coupling constant $\kappa$ for the string topogical expansion expansion is
absorbed in the
parameter $\mu$. We can re-introduce  it by the substitution $\phi \to  \kappa
^{-1/2} \phi ,
\mu\to \kappa^{-1}\mu$. Then each closed (open) string loop contributes a
factor of $\kappa^2
\ \ (\kappa )$.
 The renormalized   string tension  $\Lambda$ is  equal to   $\mu$  up to
logarithmic corrections
typical for the $D=1$ string. There are two possible critical  regimes
characterized by
different logarithmic violations (\Idis , \Imat): $ \Lambda \sim -  \mu \ln
\mu  $
  (  dilute  critical   regime) , $\Lambda \sim  \mu [\ln   \mu]^2 $  (  dense
critical   regime ). The choice of the source  stemming from the action in
\partf\ will lead to the
dense critical regime. The dilute regime is obtained by  introducing another
coupling and tuning
it.

 The second nonlinear equation \trok , which determines the open string
background, depends on the
closed   string background and    on a second parameter,
  the renormalized mass  $\mu_B\sim (\lambda^*_B -\lambda _B) \sqrt{N} $ of the
ends of the open
string.  Its general solution has been found in   \kko . Here we will restrict
ourselves to the
the case of   vanishing "quark" mass $\mu_B=0$. In this case  the solution is
given by
 \eqn\openb{\s_c= {1\over
\sqrt{2\pi}}  (  |\phi|-\sqrt \mu  )^{1/2}, \ \ \ \beta_c  + J^B =
 \sqrt{2\pi}  (|\phi|+\sqrt{\mu}
)^{1/2}.  }

\bigbreak\bigskip\bigskip\centerline{{\bf  Diagonalization of the quadratic
action }}\nobreak

 To study the string excitations we shift   the fields by their
classical values. It is also convenient to parametrize the eigenvalue interval
by the
"time-of-flight" variable $\tau$ ranging from $0$ to $\infty$
\eqn\tmof{  \tau = -\int ^{\phi}{d\phi   \over \rho_c  (\phi )} \ , \ \ \ \
\phi (\tau ) =
-\sqrt{\mu}\cosh\tau }
  and make the following redefinition of the fields
\eqn\refc{ \rho  -\rho_c =\p_{\phi} \chi = -  {\pt  \chi\over\rho_c}  , \
\ \ \ \alpha  -\alpha _c = K^-\cdot \p_{\phi} \tilde\chi = - K^-\cdot
{\pt\tilde\chi \over\rho_c} }
\eqn\refo{{ \s\over \s_c  } =  1-   { \psi\over\rho_c}  , \ \
{\beta +J^B\over \beta_c +J^B} = 1-  {\tilde\psi\over\rho_c}}
where the new fields are considered as functions of $x $ and $\tau$.
The quantum parts of the loop fields \tfdr\ are related to the fields in the
$\tau$-space by
\eqn\qpwl{W_x(\l)=\int_0^{\infty}d\tau  \ e^{-\sqrt{\mu}\cosh\tau} \pt \chi(x ,
\tau) ; \ \
\Omega_x(\l)=\int_0^{\infty}d\tau  \ e^{-\sqrt{\mu}\cosh\tau}  \psi (x ,
\tau).}

The operators $K^{\pm}$ and $K^B$ are now represented by the kernels
\eqn\kerrb{\eqalign{
\CK^+_{xx'}(\tau,\tau ') &=-C_{xx'} \pt \p_{\tau '}  \ln |\phi
+\phi ')|, \cr
\CK^-_{xx'}(\tau,\tau ') &= -2\delta_{xx'} \   \pt \p_{\tau '}   \CP \ln |\phi
-\phi '|, \cr
 \CK^B_{xx'}(\tau,\tau ') &=C_{xx'} { \s_c(\phi)  \s_c(\phi ')\over | \phi
+\phi ' |}. \cr}}
where integration is assumed to go in the interval $0<\tau<\infty$.
It is easy to see that for the  particular background \cozh , \openb ,
 in which $\phi =  - \sqrt{\mu}\cosh\tau $,
these kernels  are diagonalized by plane waves
 \eqn\orttz{\langle E, p| \tau , x \rangle = {1\over \sqrt{  \pi} }  \sin E\tau
e^{i\pi px}.}
If $x$ is considered as a continuous variable, then these kernels  represent
the following
finite-difference operators
 \eqn\kerrb{
\CK^+ = {2\pi\  \pt \cosh \p_x\over \sin \pi \pt}, \
\CK^- =  {2\pi\ \pt \cos \pi \pt\over \sin \pi \pt},\
 \CK^B = {\cosh \p_x  \over \cos \pi \pt}.}

To write the quadratic action we need the second term of the Taylor expansion
of the functional
$\CF_0$ around $\alpha_c$. This term is equal to
$\half\tilde\chi\cdot \CK^- \cdot \tilde\chi$ because the new field $\tilde
\chi$ is the fluctuating part of the spectral density in the one-matrix
integral. The quadratic action
  takes the form, in terms of the new fields\foot{The quadratic action for the
$D=0$ string theory was diagonalized  in   $\l$-space by
Moore, Seiberg and Staudacher \mss . A subtle point is that  the  eigenvalues
of the  same operator acting    in   $\l$-space and in $\tau$-space differ by a
factor $\Gamma(iE)\Gamma( -iE)$. This is possible because the two spaces are
related by a nonunitary transformation. For details see \ms . }
\eqn\zxcd{\CW^{\rm free} = \half \chi \CK^+\chi +\half \tilde\chi \CK^- (
\tilde\chi -2\chi )
+\half \psi \CK^B\psi + \half \tilde\psi ( \tilde \psi -2\psi).}
A complete diagonalization is achieved if we introduce the
ghost-like  fields
\eqn\ghostt{\chi^{(1/2)}= \tilde \chi -\chi, \ \ \ \psi ^{(1/2)}= \tilde \psi
-\psi}
 which decouple from  $\chi, \psi$,
\eqn\decuu{ \CW^{\rm free} =  \half \chi (\CK^+-  \CK^-)  \chi
  +\half \psi (\CK^B  -1) \psi + \half  \chi^{(1/2)}  \CK^-  \chi^{(1/2)}
+ \half  \psi ^{(1/2)}\psi ^{(1/2)}.}
 The  effect of the $(1/2)$-fields is that the internal  propagators are
modified by subtracting their values at $p=1/2$.  Note that the  term  to be
subtracted from the closed string propagator coincides with the loop-loop
correlator in the $D=0$ string theory.  This  can be expected, since
the expansion around the mean field (the solution of the one-matrix integral)
is in some sense expansion around the  string theory without embedding.
   It is  possible to make the subtraction  at another point but not to
eliminate it by  a redefinition of the vertices.  Without such a  subtraction
the $E$-integration would
produce  singularities when calculating loops.

We see that both  closed and open strings have the same spectrum of on-shell
states  $iE= \pm p  +2n, \ n\in \Z ,$
 that forms the light cone in a Minkowski space $(iE,p)$ with
periodic momentum  coordinate.
  Each on-shell state creates a  "microscopic loop'' on the world sheet with
given
momentum $p$ and corresponds to a local scaling operator. It can be thought of
as a product of
a vertex operator ( the state with minimal energy
$E=|p|$) and local operators representing infinitesimal
deformations of the microscopic loop.
The states with given momentum $p$ form an infinite tower of "gravitational
descendants" of this
vertex operator.

\bigbreak\bigskip\bigskip\centerline{{\bf   Interactions }}\nobreak

The interaction part of the action \fpfp\ consists of all terms that
 disappear in the planar limit. Thus we treat as interaction the nonplanar
corrections to the tadpoles and the quadratic term mixing the open and
 closed string fields.
In terms of the new fields \refc\ the interaction potential reads
 \eqn\potu{\eqalign{ \CW^{\rm int}&= \sum_x \big\{\CF_0[\alpha_c +K^-{\cdot}
\p_{\phi}\tilde \chi_x]
- \int d\tau  (\rho_c^2 -\pt\chi) \ln(1-\ \tilde\psi / \rho_c) \big\}_{<}\cr
&= \CU^{{\rm  closed}}[\tilde \chi] +\CU_{\rm open}[\tilde\psi]+\CU_{\rm
open}^{\rm  closed}
[\chi, \tilde\psi] \cr}}
 where $\{\ \ \}_{<}$ means that only the negative powers of   $\mu$ are
retained.
The individual terms in the expansion of \potu\ in the fields and in $1/\mu$
can be
associated with  surfaces with negative global curvature, localized at the
sites $x\in \Z$.

  To fix the form of the closed string vertices we  need to know the Taylor
expansion
\eqn\tayl{\CF_0[\alpha_c+\alpha_x]-\CF_0[\alpha_c ]= \sum_{n=1}^{\infty}
{1\over n!}
\int d^n\phi\ A_n \cdot \alpha_x^{\otimes n}.}
 The coefficient functions $A_n(\phi_1 , ...,\phi_n)$ are
the $n$-point correlation functions for the spectral density in  the
one-matrix integral with potential $V(\phi) = -\alpha_c(\phi)$, and can be
obtained as the
discontinuities of the $n$-loop correlators.    A closed  expression  for the
tree-level  loop
amplitudes  for an arbitrary potential was found by
  Ambjorn,   Jurkiewicz and  Makeenko
\ref\ajm{J. Ambjorn, J. Jurkiewicz and
Yu. Makeenko,  \pl 251 B (1990) 517}.
The problem is not yet completely solved but the general form of the loop
amplitudes beyond the tree
level is known \Idis , \ackm .
  We have, for the potential $V(\phi)= -\alpha_c(\phi)= (2/\pi) \tau\sinh\tau$,
 \eqn\anprp{A _n (\phi_1, ..., \phi_n ) \prod _{k=1}^n {\p \phi_k\over
\p\tau_k}
 = \mu^{ 2-n}
\Bigg[    A_n\Big({1\over\mu} ,  {\p\over \p a}\Big) \   {\p^{n-3}\over \p a
^{n-3}} \
 \prod _{k=1}^n  {\p\over \p \tau_k}\sin \Big({\p\over \p \tau_k}\Big)
 {1\over \sqrt{  \cosh \tau_k +a}   }  \Bigg]_{a=1} }
   where the function $A_n$  is  defined as the  formal series
\eqn\pnh{A_n(x,y)= \sum_{ h =  0}^{\infty} \sum _{k= 0}^{ 3h}       A^{(h,
k)}_{n }\
x^{2h}\ y^k.}
where the 3  coefficients   with $n+2h-2\le 0$  are assumed equal to zero. The
coefficient
 $A^{(h, k)}_{n }$ can be associated with a sphere with $n$ boundaries and $h$
handles.
The origin of the factors $\sin (\p/\p\tau _k)$ is that   the  correlation
functions
for the spectral density are equal to the    discontinuities of the loop
amplitudes  as functions
of the complex variables $z_k=\sqrt{\mu} \cosh \tau_k$ along the cuts $-\infty
<z_k<-\sqrt{\mu}$.
of
Using the operator representation \kerrb\   we see that  the potential \potu\
depends on the
  field $\tilde\chi$ through a discrete set of
projections $\Pi_n$  whose generating function is given by
\eqn\genn{ \Pi(a) \tilde \chi\equiv {\pi\over\sqrt{2}} \sum _n\Big( {a-1 \over
2}\Big)^n  \Pi_n \tilde  \chi = \int_0^{\infty}
   {d\tau\over   \sqrt{\cosh\tau + a} }\pt \ \cos\pi \pt \ \tilde  \chi(\tau,
x).}
By   Fourier-transforming and using the   identity
\eqn\legg{\sqrt{2} {\cosh\pi E \over \pi} \int_0^{\infty} d\tau{   \cos
E\tau\over \sqrt{\cosh\tau + a} }=
 P_{-\half +iE}(a)=    \sum _{n=0}^{\infty}
  \Big( {1-a\over
2}\Big)^n    { (\half +iE)_n  (\half -iE)_n \over n! \ n!}    }
we   find the explicit expression of \genn\  in terms of the derivatives
of  $\tilde\chi$ at the point $\tau =0$
\eqn\pipi{\Pi_n \tilde \chi  ={1\over( n!)^2}  \Bigg( \pt
\prod_{j=0}^{n-1}[(j+\half  )^2 - \pt^2]
\tilde \chi (\tau, x)\Bigg)_{\tau =0}}

 Thus   the     potential  for the closed string interactions depends on the
field  $\tilde\chi$
only through its normal derivatives  $\pt^{2n+1}\chi(\tau , x), n=0,1,...,$
along the edge $\tau =0$
 of the
half-plane
 \eqn\pocl{\CU_{{\rm  closed}} (\tilde\chi) = \sum_x
\sum _ {n=1}^{\infty}     \mu^{ 2-n}      \Bigg[
{\p^{n-3}\over \p a ^{n-3}}
 A_n\Big({1\over\mu} ,  {\p\over \p a}\Big)
     { [\Pi(a) \tilde \chi   ]^n\over n!}  \Bigg]_{a=1} }
 The   interaction  of   closed   strings   is nonzero only along the wall
$\tau =0$, which
qualitatively in
accord with    the  collective theory for strings with continuum target space
\djv , \ref\pll{J.
J. Polchinski, \np 362 (1991) 125 }.

The potential for the interaction of open strings consists of two terms. The
first term describes
interaction  involving only open strings
\eqn\ploo{\CU_{\rm open} [\tilde \psi ] =\sum _{n=3}^{\infty}
{\mu ^{1-n/2} \over n}    \sum
_x \int d\tau  [ \sinh \tau]^{2-n} [\tilde \psi (x,\tau)]^n }
  and its $n$-th term  has the geometrical meaning of a disc where $n$ strips
meet.
The factor $\mu^{1-n/2}$ is associated with the geodesic curvature
of the pieces of  boundary separating the strips.
    Contrary to the closed string, the
potential for open string interactions is smooth and only exponentially
decaying in the bulk.
The  second term
\eqn\ococ{\CU_{\rm open}^{\rm closed}=\sum _{n=1}^{\infty}
  {1  \over n}\mu ^{-n/2}   \sum _x
\int d\tau ( \sinh  \tau )^{-n}   \chi(x,\tau) \pt [\tilde \psi(x,\tau)]^n}
 describes the interaction one open string and a number of open strings. The
$n$-th term has the
geometrical meaning of a surface with the topology of a cylinder connecting one
tube and $n$ strips.
 The lowest vertex $(n=1)$ describes the transition between one open and  one
closed string
state. The nonplanar corrections to the open string tadpole are   are composed
from one  such  vertex
  and a nonplanar closed-string
tadpole, etc.

 \smallskip

In conclusion, we have constructed, up to some numerical factors,  the complete
interacting potential for the field theory of closed and open strings with
discrete target space. We believe that this potential describes as well the
interactions of closed and open  strings with target space $\R$.
 The tree-level dynamics in the open-string sector following from the potential
\ploo\ is in qualitative agreement with the  amplitudes  obtained  by
Bershadsky and Kutasov \berk\ from the Liouville theory. Moreover, their  open
string amplitudes follow from an effective lattice model very similar to our
collective theory in the $\tau$-parametrization.

Here we  considered only the case of massles fermions, $\mu_B=0$.
  If  $\mu_B\ne 0$, then the open string background is  asymptotically
approaching \openb\ when $\tau \to\infty$,   the deviation  being exponentially
small in $\tau$.
This weak dependence  on $\tau$   will affect the interaction potential but not
the   the spectrum of   the open string excitations (for  details see \kko).
     In the same way the spectrum of the
closed string does not depend on the
  string tension $\mu$.  This stability of the spectrum is the major
 discrepancy between
the  bosonic string and the strings expected to describe the dynamics of flux
tubes in QCD.

 \bigbreak\bigskip\bigskip\centerline{{\bf
Acknowledgments}}\nobreak

The author thanks the Mathematical Department of the Stockholm University and
the Elementary
Particles sector of  SISSA for hospitality during  the course of this work.
  It is a pleasure to thank  M. Douglas,
V. Kazakov, A. Jevicki,
 M. Staudacher and S. Wadia for useful discussions, and  V. Pasquier  for a
critical reading of the manuscript.

 \listrefs

 \bye